\documentclass[10pt,showpacs,preprintnumbers,amsmath,amssymb]{revtex4}
\usepackage{units}
\usepackage{bbold,euler,newcent}
\usepackage{graphicx}
\usepackage{dcolumn}
\usepackage{bm,bbm,mathtools,esvect}
\usepackage{slashed}

\begin{document}

\title{\LARGE\sc Virial theorem and generalized momentum in\\ quaternic quantum mechanics\\\vspace{5mm}}
\author{\tt\large SERGIO GIARDINO} \vspace{1cm}
\email{sergio.giardino@ufrgs.br}
\affiliation{\vspace{3mm} Departamento de Matem\'atica Pura e Aplicada, Universidade Federal do Rio Grande do Sul (UFRGS)\\
Avenida Bento Gon\c calves 9500, Caixa Postal 15080, 91501-970  Porto Alegre, RS, Brazil}

\begin{abstract}
\noindent After a review on recent results on quaternic quantum mechanics ($\mathbb{H}$QM), we present further consistency tests that
reinforce its compatibility with the  usual complex quantum mechanics ($\mathbb{C}$QM). The novel results comprises the Virial theorem, the quantum quaternic Lorentz force, the existence of a quaternic magnetic monopole and the redefinition of the expectation value.
\end{abstract}

\maketitle
\tableofcontents

\section{\sc Introduction}

After almost one century of research, quantum mechanics remains unfinished. There are open questions involving several topics, like the generality of quantum theory, the hermiticity, the entanglement, the measurement, among others. This article concerns the proposal for generalization of quantum mechanics throughout using quaternic numbers. 

We remember that quaternions ($\mathbb{H}$) are hyper-complex numbers with three imaginary units
\cite{Rocha:2013qtt}, then $q\in\mathbb{H}$ supposes that
\begin{equation}\label{i1}
 q=x_0 + x_1 i + x_2 j + x_3 k, \qquad\mbox{where}\qquad x_0,\,x_1,\,x_2,\,x_3\in\mathbb{R}\qquad\mbox{and}\qquad i^2=j^2=k^2=-1.
\end{equation}
The imaginary units $i,\,j$ and $k$ are anti-commutative (by way of example, $ij=-ji$), consequently quaternions
are non-commutative hyper-complex numbers. In symplectic notation, a quaternic number may be written in terms of complex components, where $\,q=z_0+z_1j\,$ and $\,z_0,\,z_1\in\mathbb{C}.\,$ Thus, quaternic quantum mechanics ($\mathbb{H}$QM) inquires whether we can generalize quantum mechanics by replacing the complex  wave functions with quaternic wave functions. In symplectic notation, a quaternic wave function reads
\begin{equation}\label{i1}
 \Psi=\Psi_0+\Psi_1\,j,
\end{equation}
where $\Psi_0$ and $\Psi_1$ are complex functions. 

In fact, a quaternic quantum theory was proposed by John von Neumann and Garrett Birkhoff
\cite{Birkhoff:2017kpl}, 
 wondering whether quantum mechanics admits a formulation that is independent of a numerical system. It was the fist attempt o obtain a propositional quantum logic, and another deep question of the article was: why do the wave functions are evaluated over the complex numbers? In other words, Birkhoff and von Neumann wondered  a criterion for choosing the complexes out of the four division algebras, namely the reals ($\mathbb{R}$), the complexes
($\mathbb{C}$), the quaternions ($\mathbb{H}$) and the octonions ($\mathbb{O}$), and either what we gain or what we lose with each choice.
Ernst Stueckelberg  developed real quantum mechanics ($\mathbb{R}$QM)
\cite{Stueckelberg:1960rsi,Stueckelberg:1961rsi,Stueckelberg:1961psg,Stueckelberg:1962fra}, and this formulation
is considered equivalent to $\mathbb{C}$QM. However, Stueckelberg's theory is involved; it demands
an anti-unitary operator $\,J\,$ that replaces the complex imaginary unit $\,i\,$ in Schr\"odinger equation, and this 
anti-unitary operator $\,J\,$ is specific for each anti-commuting pair of operators \cite{Stueckelberg:1960rsi}. In 
spite of these problems, $\mathbb{R}$QM is still an object of research, and plays a role within quantum information 
\cite{Mosca:2009sqs,Borish:2013rqm,Wootters:2014ret,Wootters:2012esh} and mathematical physics 
\cite{Oppio:2016pbf}, thus indicating that alternatives to complex quantum mechanics ($\mathbb{C}$QM) may be useful for understanding new physics. The next possibility is to use quaternions to build a quantum theory. The success of this enterprise is only partial, and
various unsolved questions remain. The majority of the results in $\mathbb{H}$QM concerns what we call the anti-hermitian $\mathbb{H}$QM,
and whose development has been subsumed in a book by Stephen Adler \cite{Adler:1995qqm}.
 We remember $\mathcal{A}^\dagger$ as the adjoint of the operator $\mathcal{A}$, and that an anti-hermitian operator satisfies  $\mathcal{A}^\dagger=-\mathcal{A}$. For mathematical convenience, the anti-hermitian $\mathbb{H}$QM uses an anti-hermitian Hamiltoninan operator in the Schr\"oedinger equation, while the remaining physical operators are still hermitian. Beyond this unexplained singularity of the Hamiltonian operator, the anti-hermitian formulation of $\mathbb{H}$QM presents several problems, like the ill defined
classical limit, and the exact solutions that are scarce and difficult to compare to $\mathbb{C}$QM
\cite{Davies:1989zza,Davies:1992oqq,Ducati:2001qo,Nishi:2002qd,DeLeo:2005bs,Madureira:2006qps,Ducati:2007wp,Davies:1990pm,DeLeo:2013xfa,DeLeo:2015hza,Giardino:2015iia,Sobhani:2016qdp,Procopio:2016qqq,Sobhani:2017yee,DeLeo:2019bcw,Hassanabadi:2017wrt,Hassanabadi:2017jiz}.
The  anti-hermitian solutions are difficult to interpret because
there are no simple solutions of $\mathbb{H}$QM like the free particle and the harmonic oscillator. Consequently, the exact effects of using a wave function (\ref{i1}), that has the additional degree of freedom represented by $\Psi_1$ and the constraint generated by the quaternic imaginary unit $j$, are difficult to evaluate.

However, a recent series of papers has indicated an anternative to anti-hermitian $\mathbb{H}$QM
\cite{Giardino:2016abe,Giardino:2018lem,Giardino:2018rhs,Giardino:2017yke,Mandal:2019nsf}. First of all, it is realized that non-anti-hermitian Hamiltonian may
generate probability conserving solutions. In the second article of the series \cite{Giardino:2018lem}, it has
been proven that a well-defined classical limit is attainable in $\mathbb{H}$QM without the anti-hermitian assumption. 
Finally, it has been ascertained that this non-anti-hermitian quaternic theory is defined in a real Hilbert space, where the spectral theorem holds and the time evolution is non-abelian \cite{Giardino:2018rhs}. A technique of obtaining quaternic solutions is also
proposed in \cite{Giardino:2017yke} to the case of the free particle.

In this article, we continue grounding $\mathbb{H}$QM without the anti-hermitian assumption. The general momentum operator obtained in \cite{Giardino:2016abe} has been used to obtain the quaternic Virial theorem in Section \ref{V} and a quaterninionic quantum Lorentz force in Section \ref{L}. As a by-product, we were oblidged to redefine the expectation value in Section \ref{Q}. However, before considering these results, the next section provides a summary of the results obtained in the three previous articles. 

\section{\sc quaternic quatum mechanics in real Hilbert space\label{R}}

First of all, we point out that $\mathbb{H}$QM in real space has been proven to be mathematically consistent in
\cite{Giardino:2018rhs}, where the spectral theorem has been proven and the Fourier series and the Fourier transform
have been studied. Then, let us entertain the quaternic Schr\"odinger equation 
\begin{equation}\label{r1}
\hslash\frac{\partial\Psi}{\partial t}\,i\,=\,\left[-\frac{\hslash^2}{2m}\Big(\bm\nabla- \bm{\mathcal A}\Big)^2+U\right]\Psi,\qquad
\mbox{where}\qquad\bm{\mathcal A}=\bm\alpha i+\bm\beta j,\qquad U=V +W\,j,
\end{equation}
where $\bm\alpha$ is a real vector function, $\bm\beta$ is a complex vector function and the quaternic scalar potential $U$ comprises the complex functions $V$ and $W$. We point out the position of the imaginary unit $\,i\,$ at the right hand side of $\partial_t\Psi$. The non-commutativity of $\Psi$, implies that $i\Psi\neq\Psi i$, and hence the choice $\,i\partial_t\Psi\,$ in (\ref{r1}) generates another wave equation, which we briefly discuss in the appendix.

Irrespective of that, the proposition of (\ref{r1}) demands a series of studies concerning the physical and mathematical consistency of the quantum theory described through it. First of all, the a right multiplication of (\ref{r1}) by $(-i)$ does not necessarily generate an anti-hermitian operator, and consequently the equation cannot be treated using the current anti-hermitian $\mathbb{H}$QM formalism. Of course, one could do it, but this means a restriction of the solutions, and we seek the maximal generality. Then, we firstly ascertain the interpretation of the wave function as a density of probability. Defining $\,q^*=z_0^*-z_1j\,$ as the quaternic conjugate of the arbitrary quaternion $\,q,\,$ the continuity equation \cite{Giardino:2018lem} reads
\begin{equation}\label{r2}
\frac{\partial \rho}{\partial t}+ \bm{\nabla\cdot J}=g,
\end{equation}
where $\rho=\Psi\Psi^*$ is the probability density, $g$ is a probability source, $\bm J$ is the probability current and $\bm\Pi$ is the generalized momentum operator, so that
\begin{equation}\label{r3}
g=\frac{1}{\hslash}\Big(\Psi i\,\Psi^*U^*-U\Psi i\Psi^* \Big),\;\;\;\;
\bm J=\frac{1}{2m}\Big[(\bm\Pi\Psi)\Psi^*+\Psi(\bm\Pi\Psi\big)^* \Big]\;\;\;\;\mbox{and}\qquad
\bm\Pi\Psi=-\hslash\big(\bm\nabla-\bm{\mathcal A}\big)\Psi i.
\end{equation}
(\ref{r2}) indicates that the probability density is conserved for $g=0$, and this is achieved for real $U$, in agreement with complex quantum mechanics ($\mathbb{C}$QM). In the case of complex potentials (where $W=0$), (\ref{r2}) recovers the continuity equation valid for non-hermitian $\mathbb{C}$QM \cite{Bender:2007nj,Moiseyev:2011nhq} irrespective of the non-hermitian vector potential $\bm\beta$. This coincidence permit us to formulate the hypothesis of a relation between $\mathbb{H}$QM and non-hermitian $\mathbb{C}$QM. This idea is of central importance and will be considered in future investigations.

A second element of the non-anti-hermitian $\mathbb{H}$QM comes by analogy with $\mathbb{C}$QM, where the momentum
expectation value is related to the probability current through $\langle \bm \Pi\rangle=\langle \bm J\rangle/m$. This analogy has been used in \cite{Giardino:2018lem} to define the expectation value for an arbitrary quaternic operator $\mathcal{O}$  as
\begin{equation}\label{r4}
\langle\mathcal O\rangle= \frac{1}{2}\int dx^3\Big[\big(\mathcal{O}\Psi\big)\Psi^* +\Psi\big(\mathcal{O}\Psi\big)^* \Big].
\end{equation}
The expectation value $\langle\mathcal O\rangle$ is always real, irrespective of the hermiticity of $\mathcal O$, something desirable physically. Furthermore, (\ref{r4}) recovers the $\mathbb{C}$QM expectation value for complex wave functions and hermitian operators. Conversely, (\ref{r4}) has an interesting consequence: the Hilbert space is real, in contrast to $\mathbb{C}$QM, which is developed over a complex Hilbert space.  Thus, the expression for the expectation was obtained using a physical motivation but it must satisfy several mathematical requirements that are necessary to functions that belong to a Hilbert space. The most important requirement is whether the  spectral theorem holds and  consequently establishes the correspondence between physical observables and eigenvalues. In a real Hilbert space, the fundamental theorem of algebra indicates that a restricted class of operators may have physical significance. On the other hand, in $\mathbb{C}$QM the more general complex Hilbert space is restricted to a subspace where the wave functions are physically meaningful by using hermitian operators. Such a procedure is not necessary in the case of the real Hilbert space because every eigenvalue is real an potentially meaningful. Only the application can decide the suitable formalism for each physical situation, but the most fundamental mathematical questions concerning these matters were addressed in   \cite{Giardino:2018lem,Giardino:2018rhs}, where the details may be found.

A further important consistency test is the Ehrenfest theorem, which describes the time evolution of the quantum expectation values as a classical dynamics, and that has been considered in \cite{Giardino:2018lem}. For the position expectation value, we have
\begin{equation}\label{r5}
\frac{d \langle \bm r \rangle}{dt}=\frac{\langle \bm \Pi\rangle}{m}-\frac{2}{\hbar}\big\langle (U\bm r|i)\big\rangle,
\end{equation}
where we define the notation
\begin{equation}\label{r6}
 (a|b)\Psi=a\,\Psi b.
\end{equation}
 We observe that the second term
\begin{equation}
\big\langle (U\bm r|i)\big\rangle=\int\bm r\Big(U\Psi i\Psi^*-\Psi i \Psi^* U^*\Big)dx^3
\end{equation}
is identically zero for real $U$ and pure imaginary for complex $U$. Consequently, the
dynamics of the position expectation value recovers the classical dynamics for real $U$ and recovers the $\mathbb{C}$QM for complex $U$, another exact correspondence between  $\mathbb{H}$QM and non-hermitian $\mathbb{C}$QM. Furthermore, the momentum expectation value gives
\begin{equation}\label{r7} 
\frac{d\langle p_x\rangle}{dt}=\int dx^3\left(U\Psi\frac{\partial\Psi^*}{\partial x}+\frac{\partial\Psi}{\partial x}\Psi U^*\right).
\end{equation}
For real $U$, the right hand side of (\ref{r7}) gives $\langle-\,\partial_x U\rangle$, in perfect agreement with hermitian $\mathbb{C}$QM. 
Using expectation values, we may express (\ref{r7}) as 
\begin{equation}\label{r8} 
\frac{d\langle p_x\rangle}{dt}=2\left\langle - \frac{\partial U}{\partial x}\right\rangle+
2\left\langle -U\frac{\partial}{\partial x} \right\rangle.
\end{equation}
As in the expectation value of the position, the dynamics is classical for real $U$ and recovers the $\mathbb{C}$QM when $W=0$ and 
$U$ is consequently complex. Thus, classical limit of $\mathbb{H}$QM is well defined, and the case when the dynamics is non-classical are also known. Anti-hermitian $\mathbb{H}$QM does not address these issues satisfactorily. Finally, the dynamical equation for the expectation values has been obtained in \cite{Giardino:2018lem}, so that
\begin{equation}\label{r9}
\frac{d}{dt}\Big\langle \big(\mathcal{O}|i\big) \Big\rangle
=\left\langle\frac{\partial}{\partial t}\big(\mathcal O|i\big) \right\rangle
+\frac{1}{\hslash}\left\langle \Big[\mathcal O,\,\mathcal H \Big]\right\rangle
+\frac{\partial\langle(\mathcal O|i)\rangle}{\partial t},
\end{equation}
which may also be written
\begin{equation}\label{r10}
\frac{d}{dt}\Big\langle\mathcal{O} \Big\rangle
=\left\langle\frac{\partial}{\partial t}\mathcal O \right\rangle
+\frac{1}{\hslash}\left\langle \Big[\mathcal H,\,(\mathcal O|i)\Big]\right\rangle
+\frac{\partial\langle\mathcal O\rangle}{\partial t}.
\end{equation}
If $\Psi$ is complex and $\mathcal{O}$ is complex and hermitian, (\ref{r10}) recovers the $\mathbb{C}$QM equations for the dynamics of expectation values in the Schr\"odinger picture.  Thus, there are clear evidences of a quaternic theory that contains the complex quantum mechanics as a particular case. However, there is a remaining question, whether (\ref{r9}) and (\ref{r10}) are equivalent or not. This paper considers this question in the following sections, and then let us start considering the virial theorem for $\mathbb{H}$QM.
 
\section{\sc The quaternic virial theorem\label{V}}
Let us start with another evidence of a well behaved classical limit of the $\mathbb{H}$QM: the virial theorem. 
Considering $|\bm{\mathcal A}|=0$ in the Hamiltonian of (\ref{r1}), the time evolution of the expectation values (\ref{r10}) encompasses the following stationary dynamics 
\begin{align}\label{v1}
\frac{d}{dt}\langle \bm{r\cdot p} \rangle&=\frac{1}{\hslash}\Big\langle\big[\mathcal{H},\,(\bm{r\cdot p}|i)\big]\Big\rangle\qquad
\mbox{where}\qquad \bm{p}\Psi=-\hslash\bm\nabla\Psi\,i
\end{align}
Thus, we have
\begin{align}\nonumber
 \big[\mathcal H,\,(\bm{r\cdot p}|i)\big]&=\big[\mathcal H,\,\bm{r}\big]\bm\cdot(\bm p|i)+\bm{r\cdot }\big[\mathcal H,\,(\bm{p}|i)\big]\\
\label{v2}
&=\frac{1}{2m}\big[p^2,\,\bm r\big]\bm\cdot(\bm p|i)+\bm {r\cdot} \big[U,\,(\bm p|i)\big],
\end{align}
where $\,[p^2,\,\bm p]=[U,\,\bm r]=0\,$ have been used. Using the identities
\begin{equation}\label{v3}
 \big[p^2,\,r_a]=\bm p\bm\cdot\big[\bm p,\,r_a\big]+\big[\bm p,\,r_a\big]\bm{\cdot p}
\qquad\mbox{and}\qquad \big[p_a,\,r_b\big]=-\hslash\,\big(\delta_{ab}\big|i\big)
\end{equation}
we get
\begin{align}\label{v4}
\big[p^2,\,\bm r\big]=-2\hslash\big(\bm p\big|i\big)\qquad\mbox{and}\qquad
\big[\mathcal{H},\,(\bm{r\cdot p}|i)\big]=\frac{\hslash}{m}p^2-\hslash\,\bm{r\cdot\nabla}U.
\end{align}
Consequently,
\begin{equation}\label{v5}
\frac{1}{\hslash} \left\langle\big[\mathcal{H},\,(\bm{r\cdot p}|i)\big]\right\rangle=\frac{1}{m}\langle p^2\rangle
-\frac{1}{2}\left\langle\bm{r\cdot\nabla}(U+U^*)\right\rangle
\end{equation}
In a stationary state, the time derivative on the left hand side of (\ref{v1}) is zero, and thus we obtain the virial theorem. For real $U$, the result is identical to the $\mathbb{C}$QM virial theorem. However, using a complex $U$, where $W=0$, an imaginary contribution depending on $\,V-V^*\,$ lacks on the right hand side of (\ref{v5}), the $\mathbb{C}$QM result is not recovered. This residual complex contribution is eliminated 
because the expectation value (\ref{r4}) is always real, and the pure imaginary components of $U$ do not contribute to (\ref{v5}). We may have a further insight considering equation (\ref{r9}). Firstly, 
\begin{equation}
 \big[\mathcal{H},\,\bm{r\cdot p}\big]=-\frac{\hslash}{m}\left(p^2\big|\,i\,\right)+\hslash\,\left(\bm{r\cdot\nabla}U\big|\,i\,\right),
\end{equation}
and thus
\begin{align}\nonumber
\frac{d}{dt}\big\langle (\bm{r\cdot p}|i) \big\rangle&=\frac{1}{\hslash}\Big\langle\big[\mathcal{H},\,\bm{r\cdot p}\big]\Big\rangle\\
\label{v6}
&=\frac{1}{2}\left\langle\Big(\bm{r\cdot\nabla}(U-U^*)\Big|i\Big)\right\rangle.
\end{align}
For the contributions to be correctly considered, we have to sum (\ref{v5}) and (\ref{v6}), so that
\begin{equation}\label{v7}
\frac{d}{dt}\big\langle \bm{r\cdot p} +(\bm{r\cdot p}|i)\big\rangle=\frac{1}{m}\langle p^2\rangle
-\frac{1}{2}\left\langle\bm{r\cdot\nabla}(U+U^*)\right\rangle
+\frac{1}{2}\left\langle\Big(\bm{r\cdot\nabla}(U-U^*)\Big|i\Big)\right\rangle
\end{equation}
The left hand side of (\ref{v7}) is zero because we are considering stationary states, and now we have the complete quaternic counterpart of the virial theorem, where the non-hermitian virial theorem is recovered for complex $U$.  The result is a piece of evidence that (\ref{r9}) and (\ref{r10}) are not equivalent, and the correct expression for the virial theorem must consider the sum of both of the expressions.
In the next section we give further arguments that reinforce the validity of adding (\ref{v5}) and (\ref{v6}) to obtain (\ref{v7}), and in Section \ref{Q} we propose a general expression.
\section{\sc The quaternic quantum Lorentz force\label{L}}
In order to get a quaternic version of the quantum Lorentz force, (\ref{r10}) furnishes
\begin{equation}\label{l1}
\frac{d}{dt}\Big\langle\mathcal{\bm\Pi} \Big\rangle
=\left\langle\frac{\partial}{\partial t}\bm\Pi \right\rangle
+\frac{1}{\hslash}\left\langle \Big[\mathcal H,\,(\bm\Pi|i)\Big]\right\rangle
+\frac{\partial\langle\bm\Pi\rangle}{\partial t},
\end{equation}
where
\begin{equation}\label{l2}
\Big[\mathcal H,\,(\bm\Pi|i)\Big]=\frac{1}{2m}\Big[\Pi^2,\,(\bm\Pi|i)\Big]+\Big[U,\,(\bm\Pi|i)\Big].
\end{equation}
On the other hand, from the total linear momentum $\;\Pi^2=\Pi_1^2+\Pi_2^2+\Pi_3^2\;$ and from an arbitrary 
component $\;\Pi_c\;$ of $\;\bm\Pi$, we have
\begin{eqnarray}\nonumber
 \big[\Pi^2,\,(\Pi_c|i)\big]&=&\bm{\Pi\cdot}\big[\bm\Pi,\,(\Pi_c|i)\big]+\big[\bm\Pi,\,(\Pi_c|i)\big]\bm{\cdot\Pi}\\ 
\label{l3}
&=&\Pi_a\big[\Pi_a,\,(\Pi_c|i)\big]+\big[\Pi_a,\,(\Pi_c|i)\big]\Pi_a
+\Pi_b\big[\Pi_b,\,(\Pi_c|i)\big]+\big[\Pi_b,\,(\Pi_c|i)\big]\Pi_b.
\end{eqnarray}
Introducing the notation
\begin{equation}\label{l4}
 \mathcal{P}_{[a}\mathcal{Q}_{b]}= \mathcal{P}_{a}\mathcal{Q}_{b}- \mathcal{P}_{b}\mathcal{Q}_{a},
\end{equation}
we obtain
\begin{equation}\label{l5}
 \big[\Pi_a,\,(\Pi_b|i)\big]=\hslash^2\Big(\partial_{[a}\mathcal A_{b]}-\mathcal A_{[a}\mathcal A_{b]}\big|i\Big),
\end{equation}
where
\begin{align}\label{l6}
\partial_{[a}\mathcal A_{b]}&=\;\partial_{[a}\alpha_{b]}i\,+\,\partial_{[a}\beta_{b]}\,j\;=\;\varepsilon_{abc}\Big[i\big(\bm{\nabla\times\alpha}\big)_c +\big(\bm{\nabla\times\beta}\big)_c j\,\Big],\\	
\mathcal A_{[a}\mathcal A_{b]}&=-\beta_{[a}^{\,}\beta^*_{b]}+2\alpha_{[a}\beta_{b]}ij=\varepsilon_{abc}\Big[\bm{\big(\beta^*\times\beta}\big)_c+2i\big(\bm{\alpha\times\beta}\big)_cj\,\Big],
\end{align}
and $\varepsilon_{abc}$ is the anti-symmetric Levi-Civita symbol. After defining the vectors
\begin{equation}\label{l7}
 \bm\kappa=i\,\bm{\nabla\times\alpha}\,+\,\bm{\beta\times\beta^*},\qquad
\bm\lambda=\bm{\nabla\times\beta}\,+\,2i\,\bm{\beta\times\alpha}\qquad\mbox{and}\qquad\bm{\mathcal B}=\bm\kappa+\bm\lambda j,
\end{equation}
equation (\ref{l5}) gives
\begin{equation}\label{l8}
 \big[\Pi_a,\,(\Pi_b|i)\big]=\hslash^2\varepsilon_{abc}\big(\,\mathcal B_c\,\big|\,i\,\big).
\end{equation}
We call $\bm{\mathcal B}$ the quaternic magnetic field, and it is a pure imaginary quaternic vector, so that $\,\bm{\mathcal B}^*=-\bm{\mathcal B}.\,$ In order to handle the quaternic vector $\bm{\mathcal B}$, one defines a vector product between the quaternic vectors $\;\bm X=\bm X_0+\bm X_1j\;$ and 
$\;\bm Y=\bm Y_0+\bm Y_1j,\;$ namely
\begin{equation}\label{l9}
 \bm X\times\bm Y\,=\, \bm X_0\times\bm Y_0\,-\,\bm X_1\times\bm Y_1^*\,+\,\big(\bm X_0\times\bm Y_1\,+\,\bm X_1\times\bm Y_0^*\big)\,j.
\end{equation}
This quaternic vector product is not identical to the usual real or complex vector product, one immediately sees that $\,\bm{X\times Y}\neq -\bm{Y\times X}.\,$ Next, using (\ref{l3}), (\ref{l8}) and (\ref{l9}), we get
\begin{align}\label{l10}
\big[\,\Pi^2,\,(\bm\Pi|i)\,\big]=\hslash^2\Big[\big(\bm{\mathcal B}\big|i\big)\times\bm\Pi\,-\,\bm\Pi\times\big(\bm{\mathcal B}\big|i\big)\Big]
\end{align}
However, $\langle\bm{\nabla\times\mathcal{B}}\rangle=0$ implies that
\begin{align}\label{l11}
\frac{1}{\hslash^3} \left\langle\Big[\Pi^2,\,(\bm\Pi|i)\Big]\right\rangle=&\;\big\langle\bm{\mathcal A\times \mathcal B}\big\rangle
-\big\langle\bm{\mathcal B\times \mathcal A}\big\rangle\\
\nonumber
=&\,-\,2\Big\langle\bm{\alpha\times\nabla\times\alpha}\,+\,
\bm{\beta\times\nabla\times\beta}^*\,-\,4i\,\bm{\alpha\times(\beta\times\beta^*)}\big\rangle,
\end{align}
and the second commutator term of (\ref{l2}) additionally renders
\begin{equation}\label{l12}
 \big[\,U,(\bm\Pi|i)\big]=\hslash\big[\,\bm{\mathcal A},\,U\big]-\hslash\bm\nabla U\qquad \mbox{and}\qquad
\Big\langle\big[\,U,\,(\bm\Pi|i)\big]\Big\rangle=-\frac{\hslash}{2}\Big\langle\bm\nabla \big(U+U^*\big)\Big\rangle.
\end{equation}
Consequently (\ref{l1}) allots
\begin{equation}\label{l13}
\frac{d}{dt}\Big\langle\mathcal{\bm\Pi} \Big\rangle
=\hslash^2\Big[\big\langle\bm{\mathcal A\times \mathcal B}\big\rangle-
\big\langle\bm{\mathcal B\times \mathcal A}\big\rangle\Big]-\frac{1}{2}\Big\langle\bm\nabla \big(U+U^*\big)\Big\rangle+
\left\langle\frac{\partial(\bm{\mathcal{A}}|i)}{\partial t}\right\rangle,
\end{equation}
where it has been used that
\begin{equation}\label{l14}
 \left\langle\frac{\partial\bm\Pi}{\partial t}\right\rangle=\hslash\left\langle\frac{\partial(\bm{\mathcal{A}}|i)}{\partial t}\right\rangle.
\end{equation}
Equation (\ref{l13}) neither describes a classical dynamics nor recovers the $\mathbb{C}$QM result in a complex limit. It is physically meaningless. However, as has been done for the virial theorem, let us then consider (\ref{r9}) with $\mathcal O=\bm\Pi$. 
In this case, we have
\begin{align}\label{l15}
\frac{d\langle(\Pi|i)\rangle}{dt}=&\;\hslash\,\big\langle\bm{\mathcal B}\times\bm p-\bm p\times\bm{\mathcal B}\big\rangle
+\frac{1}{2}\Big\langle\bm\nabla \Big(U-U^*\Big|i\Big)\Big\rangle+\left\langle\big[U,\,(\bm{\mathcal A}|i)\big]\right\rangle,
\end{align}
where it has been used that
\begin{equation}
 \left\langle\frac{\partial(\bm\Pi|i)}{\partial t}\right\rangle=-\hslash\left\langle\frac{\partial\bm{\mathcal{A}}}{\partial t}\right\rangle=0,
\end{equation}
and also that
\begin{equation}\label{l16}
 \big[\,U,\bm\Pi\big]=\hslash\big[U,\,(\bm{\mathcal A}|i)\big]+\hslash(\bm\nabla U|i)\qquad \mbox{and}\qquad
\left\langle\big[\,U,\bm\Pi\big]\right\rangle=\hslash\left\langle\big[U,\,(\bm{\mathcal A}|i)\big]\right\rangle+
\frac{\hslash}{2}\Big\langle\bm\nabla \Big(U-U^*\Big|i\Big)\Big\rangle.
\end{equation}
In the same fashion as has been done for the virial theorem, we sum (\ref{l13}) and (\ref{l15}) in order to obtain a quaternic
quantum Lorentz force 
\begin{align}\label{l17}
\Big\langle\bm F_{\,\mathbb H}\Big\rangle=& \frac{d\langle \Pi\rangle}{dt}+\frac{d\langle(\Pi|i)\rangle}{dt}=\\\nonumber
=&\;\hslash\,\big\langle\bm{\mathcal B}\times\bm p-\bm p\times\bm{\mathcal B}\big\rangle
+\hslash^2\Big\langle\bm{\mathcal A\times \mathcal B}-
\bm{\mathcal B\times \mathcal A}\Big\rangle+\left\langle \frac{\partial(\bm{\mathcal A}|i)}{\partial t}\right\rangle-
\,\left\langle\bm\nabla\frac{U+U^*}{2}\right\rangle
+\left\langle\bm\nabla \frac{U-U^*}{2}\Big|i\right\rangle+\left\langle\big[U,\,(\bm{\mathcal A}|i)\big]\right\rangle.
\end{align}
The above result does make sense: it recovers the $\mathbb{C}$QM result in the complex limit and there are contributions for each of the potentials. At the present point, it is too early to know precise the nature of the new quaternic potential and fields $\,\bm{\mathcal A},\,\bm{\mathcal B}\,$ and the imaginary components of $U$, but the existence of a dynamical equation is the first step to ascertain their physical character. As has been said in the beginning of this article, ascertaining the physical meaning of these fields is an exciting challenge for future research. However, we may attain an initial conclusion from the results. The definition of the quaternic magnetic field (\ref{l7}) immediately gives
\begin{equation}
 \bm{\mathcal B}=\bm{\nabla\times\mathcal A}-\bm{\mathcal{A}\times\mathcal{A}}\qquad\mbox{where}\qquad\bm{\mathcal{A}\times\mathcal{A}}=\bm{\beta\times\beta}^*+2i\,\bm{\beta\times\alpha}\,j.
\end{equation}
Consequently
\begin{equation}\label{l17}
\langle \bm{\nabla\cdot\mathcal B}\rangle= 0,\qquad\qquad\mbox{but}\qquad\qquad \langle (\bm{\nabla\cdot\mathcal B}|i)\rangle\neq 0.
\end{equation}
Hence, the model naturally predicts the existence of a quantum magnetic monopole, something that is not predicted in $\mathbb{C}$QM. Relations between quaternions and magnetic monopoles are already known \cite{Hitchin:1988dmm}, and this relation indicates that the result is correct as much as indicates an
exciting direction for future research.
\section{\sc quaternic quantum formalism revisited\label{Q}}
 In Section \ref{R} an ambiguity in the dynamical evolution of the expectation value was observed in (\ref{r9}) and (\ref{r10}). In Section \ref{V} and in Section \ref{L}  we obtained that the sum of (\ref{r9}) and (\ref{r10}) give physically consistent results for the momentum of a particle. Consequently, the results of the preceding sections enable us to revisit the introduction and emend the formalism of $\mathbb{H}$QM. A novel expectation value have to be defined, namely
\begin{equation}\label{q1}
\big\langle\mathcal O_\mathbb{H}\big\rangle=\big\langle\,\mathcal O +\big(\mathcal O\,|\,i\big)\big\rangle
=\big\langle \big(\mathcal O\,|\,1+i\big)\big\rangle.
\end{equation}
The expectation value (\ref{q1}) resembles a complexification of the real Hilbert space, something already considered in \cite{Sharma:1988css,Sharma:1988chs}. In fact, these works argue that real Hilbert spaces must be ruled out of quantum mechanics, while other works point out the fundamental role of complex
numbers in quantum mechanics \cite{Toyoda:1973nch,Lahti:2017wch}. Furthermore, the formulation of anti-hermitian
$\mathbb{H}$QM in a quaternic Hilbert space is also criticized \cite{Graydon:2013sra,Gantner:2017ech}. We understand that our results are in agreement with these previous results. (\ref{q1}) couples a complex structure to a quaternic operator, where $\mathcal{O}\to(\mathcal{O}|1+i)$ determines a physical measure, thus imposing a difference between the inner product and the physical expectation value. This conciliation between a real Hilbert space and complex numbers must be confirmed in future research.

Further evidences in favour of (\ref{q1}) are
the invariance of the quaternic Ehrenfest theorem, because $\langle (x|i)\rangle=\langle (p_x|i)\rangle=0$, and also the unification of (\ref{r9}-\ref{r10}) on a single dynamical equation for expectation values in the Schr\"odinger picture
\begin{equation}\label{i110}
\frac{d}{dt}\Big\langle\mathcal{O} +(\mathcal O|i)\Big\rangle
=\left\langle\frac{\partial}{\partial t}\Big(\mathcal O +(\mathcal O|i) \Big)\right\rangle
+\frac{1}{\hslash}\left\langle \Big[\mathcal O-(\mathcal O|i),\,\mathcal H\Big]\right\rangle
+\frac{\partial\langle\mathcal O+(\mathcal O|i)\rangle}{\partial t}.
\end{equation}
The new definition and dynamics for quantum expectation values based on physical grounds reinforce the consistency
of $\mathbb{H}$QM. It is another element that compose the whole picture of the theory and that adds on well defined wave equation, expectation values, complex limits, classical limits, and spectral decomposition. We can consider it as a solution that seeks a problem, and hence the research of exact solutions seems to be an urgent direction for future research.

\section{\sc Conclusion\label{C}}
In this article we have revisited and emended the mathematical machinery of $\mathbb{H}$QM in real Hilbert space
\cite{Giardino:2016abe,Giardino:2018lem,Giardino:2018rhs}. The previous results demonstrate that the theory is provided with  wave equation, momentum operator, conservation of probability, expectation values, classical limit and spectral decomposition. The further results of the validity of the virial theorem and of the existence of the time evolution for the generalized linear momentum provide further evidences of the consistency of the theory. 

The existence of a consistent framework is a fact that permit us formulate further questions about quaternic quantum mechanics. A general question about $\mathbb{H}$QM involves its relation to non-hermitian $\mathbb{C}$QM \cite{Bender:2007nj,Moiseyev:2011nhq}, a framework where parity and time-reversal invariances ($\mathcal{PT}-$invariances) replace hermiticity. This approach rendered fruitful theoretical developments 
\cite{Scolarici:2003wu,Scolarici:2009zz,Solombrino:2002vk,Mostafazadeh:2001jk,Mathur:2010nhh,Mathur:2014rnh}
and also experimental tests of the theory have been done 
\cite{Makris:2008dyn,Guo:2009obs,Rueter:2010pst,Regensburger:2012phl}. Inquiring whether the quaternic and the complex formulations relate themselves is very relevant. 
Conversely, specific quaternic solutions are needed to evidence the physical meaning of the pure quaternic terms of the wave function, of the scalar potential and of the gauge field $\bm\beta$. Inquiring about a quaternic magnetic monopole is another fascinating question. In fact, the research of exact solutions is the most important trend in $\mathbb{H}$QM, a task that is easier in the present situation of consistency of the theory presented in this paper.

\appendix
\section{\sc the left complex wave equation}
In this section Virial theorem and the quantum Lorentz force are calculated considering the left-complex quaternic Schr\"odinger equation, 
namely
\begin{equation}\label{a1}
i\hslash\frac{\partial\Psi}{\partial t}\,=\mathcal{H}\Psi\qquad\mbox{and}\qquad
\mathcal{H}=\frac{\hbar^2}{2m}i\Big(\bm\nabla -\bm A\Big)\bm\cdot i\Big(\bm\nabla -\bm A\Big)+U
\end{equation}
where the quaternic vector potential $\,\bm{\mathcal{A}}\,$ and scalar potential $\,U\,$ are defined in (\ref{r1}). There are 
a slight variation in the terms of the continuity equation (\ref{r4}), where the source $g$,  
the probability density $\bm J$, and the gauge-invariant quaternic linear 
momentum operator is $\bm\Pi$ are as follows
\begin{equation}\label{a2}
\rho=\Psi^*\Psi,\qquad g=\Psi^*\left(\frac{V^* i-i\, V}{\hbar}\right)\Psi,\qquad
\bm J=\frac{1}{2m}\left[\Psi^*\big(\bm\Pi\Psi\big)+\big(\bm\Pi\Psi\big)^*\Psi\,\right],
\qquad\mbox{and}\qquad\bm\Pi\Psi=-i\,\hbar\big(\bm\nabla-\bm Q\big)\Psi.
\end{equation}
The Ehrenfest theorem has slight differences in the dynamics of position and linear momentum (\ref{r5}-\ref{r8}) as well, 
which may be obtained in \cite{Giardino:2018lem}, but the conclusions are identical. 
Nevertheless, there are many differences in the expectation value dynamics between (\ref{r9}-\ref{r10}) and the dynamical equations obtained from (\ref{a1}), accordingly
\begin{align}
& \frac{d}{dt}\Big\langle \mathcal O - i\mathcal O i\Big\rangle
=\left\langle\frac{\partial}{\partial t}(\mathcal O - i\mathcal O i)\right\rangle
+\frac{1}{\hslash}\left\langle \Big[\mathcal H,\,\mathcal O i+i\mathcal O\Big]\right\rangle+
\frac{\partial}{\partial t}\Big\langle \mathcal O - i\mathcal O i\Big\rangle,\label{a3}
\\
& \frac{d}{dt}\Big\langle \mathcal O i + i\mathcal O \Big\rangle
=\left\langle\frac{\partial}{\partial t}(\mathcal O i +i\mathcal O )\right\rangle
-\frac{1}{\hslash}\left\langle \Big[\mathcal H,\,\mathcal O -i\mathcal O i\Big]\right\rangle+
\frac{\partial}{\partial t}\Big\langle \mathcal O i+ i\mathcal O \Big\rangle,\label{a4}
\\
& \frac{d}{dt}\Big\langle \mathcal O + i\mathcal O i\Big\rangle
=\left\langle\frac{\partial}{\partial t}(\mathcal O + i\mathcal O i)\right\rangle
-\frac{1}{\hslash}\left\langle \Big\{\mathcal H,\,\mathcal O i-i\mathcal O\Big\}\right\rangle+
\frac{\partial}{\partial t}\Big\langle \mathcal O + i\mathcal O i\Big\rangle,\label{a5}
\\
& \frac{d}{dt}\Big\langle \mathcal O i-i\mathcal O \Big\rangle
=\left\langle\frac{\partial}{\partial t}(\mathcal O i - i\mathcal O )\right\rangle
+\frac{1}{\hslash}\left\langle \Big\{\mathcal H,\,\mathcal O +i\mathcal O i\Big\}\right\rangle+
\frac{\partial}{\partial t}\Big\langle \mathcal O i- i\mathcal O \Big\rangle.\label{a6}
\end{align}
In the case of the virial theorem, we already have a difference between the results, so that (\ref{v7})
turns into
\begin{equation}\label{a7}
\frac{d}{dt}\big\langle \bm{r\cdot p} +(\bm{r\cdot p}|i)\big\rangle=\frac{1}{m}\langle p^2\rangle
-\frac{1}{2}\left\langle\bm{r\cdot\nabla}(U+U^*)\right\rangle
+\frac{1}{2}\left\langle\bm{r\cdot\nabla}(iU-U^*i)\right\rangle+\big\langle 2W\bm{r\cdot p}\big\rangle,
\end{equation}
The last term of (\ref{a7}) is absent in (\ref{v7}). This result confirms that the left-complex quaternic
Schr\"odinger equation (\ref{a1}) is not equivalent to (\ref{r1}). The calculation of a Lorentz force from (\ref{a1}) reinforces this conclusion. The commutators
\begin{eqnarray}
 \big[\;\Pi_a,\,\Pi_b\;\big]&=&\hslash^2\Big[\partial_{[a}A_{b]}-\Big(A_{[a}+iA_{[a}i\Big)\partial_{b]}+iA_{[a}iA_{b]}\Big]\\
 \big[\,\Pi_a,\,\Pi_b i\,\big]&=&\hslash^2\Big[\partial_aA_bi-\partial_biA_a+\Big(A_bi-iA_b\Big)\partial_a+iA_aiA_bi+iA_bA_a\Big]\\
 \big[\,\Pi_a,\,i\Pi_b\,\big]&=&\hslash^2\Big[i\partial_{[a}A_{b]}-\Big(A_bi-iA_b\Big)\partial_a-iA_aA_b+A_biA_a\Big]\\
 \big[\Pi_a,\,i\Pi_b i\big]&=&\hslash^2 \Big[i\partial_aA_bi+\partial_bA_a+\Big(A_{(a}+iA_{(a}i\Big)\partial_{b)}-iA_aA_bi-A_bA_a\Big],
\end{eqnarray}
do not seem to enable a quantum Lorentz force. We cannot say that is impossible to get such a result. 
It is of course an open question, however, the Virial theorem (\ref{a7}) leads us to expect additional terms
the Lorentz force, thence the physical interpretation  is probably more difficult, because of this, we consider 
that this result indicates that only (\ref{r1}) is physically meaningful. The wave equation (\ref{a1}) with $|\bm{\mathcal{A}}|=0$ has been adopted in anti-hermitian $\mathbb{H}$QM \cite{Adler:1995qqm}, and this fact maybe explains the difficulty in obtaining physically meaningful solutions in such framework.


%
%
%
%

\bibliographystyle{unsrt} 
\bibliography{bib_virial}

\end{document}